\documentstyle[psfig,11pt]{article}

\begin{document}

\title{Self lensing effects for compact stars and their mass-radius relation}

\maketitle

\author{A. R. Prasanna$^1$ \& Subharthi Ray$^2$\\
$^1$Physical Research Laboratory, Navrangpura,
Ahmedabad 380 009, India; e-mail: prasanna@prl.ernet.in\\
$^2$Inter University Centre for Astronomy and Astrophysics, Post
Bag 4, Pune 411 007, India; e-mail: sray@iucaa.ernet.in}

\begin{abstract}
{During the last couple of years astronomers and astrophysicists
have been debating on the fact whether the so called `strange
stars' - stars made up of strange quark matter, have been
discovered with the candidates like SAX~J1808.4$-$3658,
4U~1728$-$34, RX~J1856.5$-$3754, etc. The main contention has
been the estimation of radius of the star for an assumed mass of
$\sim1.4 M_\odot$ and to see whether the point overlaps with the
graphs for the neutron star equation of state or whether it goes
to the region of stars made of strange matter equation of state.
Using the well established formulae from general relativity for
the gravitational redshift and the `lensing effect' due to
bending of photon trajectories, we, in this letter, relate the
parameters M and R with the observable parameters, the redshift
$z$ and the radiation radius R$_\infty$, thus constraining both M
and R for specific ranges, without any other arbitrariness. With
the required inputs from observations, one ought to incorporate
the effects of self lensing of the compact stars which has been
otherwise ignored in all of the estimations done so far.
Nonetheless, these effect of self lensing makes a marked
difference and constraints on the M-R relation.}
\end{abstract}


With the new observations coming in from the satellites Chandra
and XMM Newton, it is getting trickier to decide whether many of
the X-ray emitting stars observed are just neutron stars or more
exotic stars made up of strange quark matter \cite{d98, li99a,
li99b}. The equilibrium configuration for stars with matter other
than just nucleons, need a very compact structure and thus with
the masses of around $1\sim1.5 M_{\odot}$, they have to be far
more smaller and compact than conventional neutron stars. There
have been several discussions on strange stars but the community
of astrophysicists do not seem to have any final say on this,
with several groups differing in opinion on same candidates.
However, it is clear that in spite of difficulties involved,
astronomers are being more and more successful in getting lot
more details for many candidates, particularly for the ones
having X-ray emission. Apart from getting into debates over the
issue, it is perhaps quite necessary to consider the theoretical
constraints that the presently accepted physical theories, yield
purely from logical reasons. One needs to consider basically the
mass-radius relation which arises from different possible effects
that the observations are constrained with.

As general relativity has been accepted as the most successful
theory to describe gravity, it is necessary to consider seriously
the effects it brings in while estimating parameters of the
stellar structure. One of the constraint which everyone seem to
accept gracefully is the gravitational redshift factor, while
observing a distant massive object, as most of the discussions
relate the so called {\it observed radius} $R_\infty$ (Radiation
radius) with the actual radius R, as given by

\begin{equation}
R_\infty = R A^{-1}
\label{eq:rinftyr}
\end{equation}
$A$ being the redshift factor, given by $A=(1-2m/R)^{1/2}$ for the
Schwarzschild metric representing the field of a static star,
while $A=(1-2m/R-\omega^2R^2Sin^2\theta)^{1/2}$ for the linearised
Hartle-Thorne metric \cite{ht6869}, representing the field of a
slowly rotating star with angular velocity $\omega,$ which is
given by
\begin{eqnarray}
\nonumber ds^2=(1-\frac{2m}{r})dt^2 - (1-\frac{2m}{r})^{-1}dr^2 -
r^2d\theta^2\\
- r^2Sin^2\theta(d\phi-\omega dt)^2
\label{eq:metric}
\end{eqnarray}
where $m=GM/c^2$ represents the mass M in geometrical units.
Equation (\ref{eq:rinftyr}) thus yields for nonrotating star, the
equation
\begin{equation}
R^3-R_\infty^2(R-2m)=0
\end{equation}
and the equation
\begin{equation}
R^6-R_\infty^2(R^4-2mR^3-4J^2Sin^2\theta)=0
\end{equation}
for the rotating star, where $J$ is the specific angular momentum
related to $\omega$, through the relation $\omega=2J/R^3$. With
these two, one can easily workout the actual radius R of the star
in either case, for a given $M$ and $R_\infty$ the observed
radiation radius.

\begin{figure}
\centerline{\psfig{figure=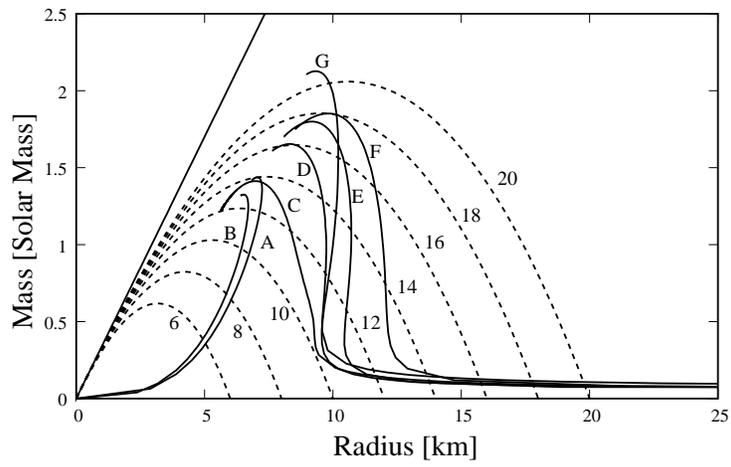,width=10cm}} \caption{The
dashed curves, labeled as 6, 8, 10, etc. (up to 20, in units of
km), show the allowed variation of mass with the radius for
different values of the radiation radius. The two lines labeled A
\& B are from the strange star equation of states. The curve
labeled C is for the hyperon star and the curves labeled D, E, F
and G are for different neutron star equation of states. The
straight line is the line for the event horizon. }
\label{fig:lens}
\end{figure}

However, it is to be remembered that general relativity predicts
another important effect, that of gravitational lensing,
associated with bending of light rays passing across the star. In
the case of extremely compact objects, the gravitational potential
M/R is so large that there could be self lensing effect, where
the radiation coming from the near neighbourhood of a compact
star would get lensed such that the star from a distance appear
much bigger than what it actually is. Nollert et al. \cite{nol89}
considered this effect in analyzing the `relativistic looks' of a
neutron star and have graphically depicted the consequences of the
relativistic light deflection. As they point out if
$I_{\nu_\infty}$ is the observed specific intensity by an
asymptotic observer, then it is related to the intrinsic
intensity $I_{\nu_s}$, through the relation
\begin{equation}
I_{\nu_\infty}=I_{\nu_s} \nu_\infty^3/\nu_s^3 = I_{\nu_s}A^3
\end{equation}
where $A$ is again the redshift factor given earlier. As $I_\nu
\propto F/R^2$, $F$ being the total flux of radiation, one can
easily find that, when self lensing effect is taken into account,
the two radii $R_\infty$ and R are related through the equation
\begin{equation}
R_\infty^2=R^2A^{-3}
\label{eq:lens2}
\end{equation}
It is to be noted that there can be some variation in the
observed flux, but considering an equilibrium state of the star,
it can be taken to be constant for a smaller time scale (of
observation). Using the expression for $A$, the relation between
$R_\infty$ and R turns out to be
\begin{equation}
R^7-R_\infty^4(R^3-6mR^2+12m^2R-8m^3)=0
\label{eq:lens}
\end{equation}
for the non rotating case and
\begin{eqnarray}
\nonumber R^{10}-R_\infty^4(R^6-6mR^5&+&12m^2R^4+
8m^3R^3\\&-&12J^2Sin^2\theta(R-2m)^2)=0
\end{eqnarray}
for the rotating case, wherein the powers of $J^4$ are neglected.
With these equations it is clear that given a $R_\infty$ and $J$,
there is no guarantee that real positive roots exist for $R$ for
any $M$. Fig. (\ref{fig:lens}) shows the plots of $M$ vs $R$ for
different values of $R_\infty$ for the case with rotation
($j=0.3m^2$). The plots for the case $j=0$ hardly differs from
these curves as the effects appear only with $j^2$ order.

Another candidate which can be taken as an example is the compact
star PSR~B0656$+$14 \cite{btgg03}. It has been studied in various
wavelengths, from optical \cite{kop01} to x-rays \cite{ms02}. Two
component blackbody models are commonly used in measuring the
surface temperature of the star. The hard component of the
blackbody spectrum is identified with the hot polar caps and the
soft component is due to the general photospheric radiation and is
directly related to the stellar surface temperature. The
redshifted surface temperature (T$_\infty$) is related to the
temperature of the blackbody fit by
\begin{equation}
T_\infty=\frac{T}{1+z}
\end{equation}
From ROSAT data, Koptsevich et al.\cite{kop01} estimated a value
of T$_\infty \sim 8 \times 10^5$K  and quite independently
Marshall and Schultz \cite{ms02}, from Chandra observations gave
similar results. This value of T$_\infty$ gives in turn, the
radiation radius (R$_\infty$) of the stellar photosphere for the
assumed distance estimate of 288 pc. Said so, with this value,
R$_\infty$ is calculated to be $\sim$ 8 km. This immediately
points the source to be a very compact strange star. So, the
believers of neutron star models immediately {\it corrected}
themselves, saying that the star's temperature has been
overestimated, and hence should be changed to a lower value in
order to give a larger value of R$_\infty$, that can go very well
with the normal neutron stars. These types of guesses are always
made to keep a star in the conventional neutron star regime.
However, none of the calculations make use of the lensing effect
that is more concrete. Even with a large range of guess values
for the surface temperature, and subsequently a large range of
values in R$_\infty$, the effect of lensing (Eq. \ref{eq:lens})
allows a smaller window for the M-R curve, than that without
lensing taken into consideration.

Recently there have been attempts to measure redshift of spectral
lines emitted from regions close to the stellar surface
\cite{cot02, san02} and using this information they obtain the
radius $R$ for an assumed mass $M$. However, if both the
observational data regarding the radiation radius $R_\infty$ and
the redshift $z$ are to be taken seriously, then one can work out
both the radius $R$ and mass $M$ of the star uniquely using
equation (\ref{eq:lens2}) and the relation
\begin{equation}
1+z=(1-2m/R)^{-1/2}
\end{equation}
as given by
\begin{equation}
R=(1+z)^{-3/2}R_\infty
\end{equation}
and
\begin{equation}
m=\frac{z(z+2)}{2(1+z)^{7/2}}R_\infty.
\end{equation}

Fig. (\ref{fig:lens}) gives the plots of $m(R)$ for given
$R_\infty$ values as well as the plots for various theoretical
compact star models made from various equation of state of the
matter. From the plot it is clear that for a given value of
$R_\infty$, there is a maximum limit of the mass up to which the
stars can exist with real mass-radius relation. The straight line
represents the event horizon for each mass, which is a natural
bound for the star not to be a black hole. Prasanna and Ray
\cite{pr04} had considered earlier this case for the star
RX~J1856.5$-$3754 \cite{dra02, wl02} which had created a lot of
controversies regarding its nature, with the estimated radiation
radius for this X-ray star being highly uncertain, ranging from
8km to 15km. It is now obvious from the figure that for these
values of $R_\infty$ the maximum allowed mass range is between
0.8 and 1.5 $M_\odot$. As one does not have the redshift
measurement for this, one can only guess and as the overlapping
curves show, the star could be either a neutron star or a strange
star. Hence it is very clear that with the existing observational
evidence, one cannot rule out the possibility of the star
RX~J1856.5$-$3754 being a strange star. However, one would say
that it is premature to come to conclusions one way or the other
unless one has both the photometric (for the radiation radius)
and the spectrometric (for the redshift) observations for these
compact objects.

This effect of lensing is nothing new, and has been discussed in
many context, but has been ignored in all the calculations for
evaluating the M-R relations of the compact stars. However, as we
showed, with this effect taken into account, they impose more
constraints in the M-R window, and hence can give a better
picture of the compact stellar candidates.


\begin{thebibliography}{}
\bibitem{d98} M. Dey, I. Bombaci, J. Dey, S. Ray \& B. C. Samanta,
Phys. Lett. B438, (1998) 123; erratum B467, (1999) 303.

\bibitem{li99a} X-D. Li, I. Bombaci, M. Dey, J. Dey \& E. P. J. van den
Heuvel, Phys. Rev. Lett., 83, (1999a) 3776.

\bibitem{li99b} X-D. Li, S. Ray, J. Dey, M. Dey \& I. Bombaci,
ApJ, 527, (1999b) L51.

\bibitem{ht6869} J. B. Hartle \& K. S. Thorne, ApJ,
153, (1968) 807; ApJ, 158, (1969) 719.

\bibitem{nol89} H. P. Nollert, H. Ruder, H. Herold \&
U. Kraus, A\&A, 208, (1989) 153.

\bibitem{cot02} J. Cottam, F. Paerels \& M. Mendez,
Nature, 420, (2002) 51.

\bibitem{san02} D. Sanwal, G. G. Pavlov, V. E. Zavlin \&
M. A. Teter, ApJ, 574, (2002) L61.

\bibitem{pr04} A. R. Prasanna \& S. Ray, ``Strange stars, have they
been discovered",  Nova Publications, (in press) (2004).

\bibitem{dra02} J. J. Drake $et~al.$, ApJ, 572, (2002) 996.

\bibitem{wl02} F. M. Walter \& J. M. Lattimer, ApJ, 576, (2002) L145.

\bibitem{btgg03} W. F. Brisken, S. E. Thorsett, A. Golden \& W. M.
Goss, ApJ, 593, (2003) L89.

\bibitem{kop01} A. B. Koptsevich $et~al.$, A\&A, 370, (2001) 1004.

\bibitem{ms02} H. L. Marshall \& N. S. Schultz, ApJ, 574, (2002)
377.



\end{thebibliography}
\end{document}